\documentclass[twocolumn]{autart}
\usepackage{amssymb}
\usepackage{algorithm}
\usepackage{algpseudocode}
\usepackage{amsmath}
\usepackage{mathrsfs}
\usepackage{graphicx}
\usepackage{float}
\usepackage{cite}
\usepackage{epstopdf}
\newcommand{\squeezeup}{\vspace{-0.75cm}}
\begin{document} 
\begin{frontmatter}
\title{Analysis of higher order time delay systems using Lambert W function}
\author[Delhi]{Niraj Choudhary}\ead{niraj.choudhary@ee.iitd.ac.in},    
\author[Delhi]{Janardhanan Sivaramakrishnan}\ead{janas@ee.iitd.ac.in},               
\author[Delhi]{Indra Narayan Kar}\ead{ink@ee.iitd.ac.in}  
\address[Delhi]{Department of Electrical Engineering\\
Indian Institute of Technology\\
Delhi, India 110016}  
\begin{abstract}
In this note, analysis of time delay systems using Lambert W function approach is reassessed. A common canonical form of time delay systems is defined. We extended the recent results of \cite{Rudy} for second order into \textit{n}th order system. The eigenvalues of a time delay system are either real or complex conjugate pairs and therefore, the whole eigenspectrum can be associated with only two real branches of the Lambert W function. A new class of time delay systems is characterized to extend the applicability of the above said method. A state variable transformation is used to transform the proposed class of systems into the common canonical form. Moreover, this approach has been exploited to design a controller which places a subset of eigenvalues at desired locations. Stability is analyesed by the help of Nyquist plot. The approach is validated through an example.
\end{abstract}

\end{frontmatter}

\section{Introduction}
A time delay system (TDS) is represented as
\begin{equation}\label{eq1}
\dot{x}(t)=Ax(t)+A_d x(t-h),
\end{equation}
where $A$ is the system matrix, $A_d$ is delayed system matrix  and $x(t)$ is an $n\times 1$ state vector. The charactristic equation of system (\ref{eq1}) is
\squeezeup\\
\begin{align}\label{eq2}
	(S-A-A_de^{-Sh})=0,
\end{align}
where $S\in\mathbb{C}^{n\times n}$. An auxiliary matrix $P$ is introduced, such that
\begin{equation}\label{eq3}
	h(S-A)e^{(S-A)h}=A_dhP,
\end{equation}
Define $M_k=hA_dP_k$. Using (\ref{eq3}), the solution matrix $S_k$ is obtained as
\vspace{-0.5cm}
\begin{equation}\label{eq4}
	S_k=\frac{1}{h}W_k(M_k)+A,
\end{equation}
where $W_k(M_k)$ is the Lambert W function of matrix $M_k$, for $k=0,\pm1,\pm2,\ldots\pm\infty$. Substituting (\ref{eq4}) into (\ref{eq2}), yields the following non-linear equation from which unknown matrix $M_k$ is obtained
\begin{equation}\label{eq5}
	W_k(M_k)e^{W_k(M_k)+Ah}=A_dh.
\end{equation}
The eigenspectrum of (\ref{eq1}) is computed by solving the following steps for branch index $k=0,\pm1,\pm2,\ldots\pm\infty$ \cite{Worldscientific}.
\begin{itemize}
\item Solve the non-linear equation (\ref{eq5}) in each relevant case.
\item Compute $S_k$ by substituting $M_k$ in (\ref{eq4}).
\item Compute the eigenvalues of $S_k$.
\end{itemize} 
The method presented above for the analysis of linear TDS using Lambert W function has been given in \cite{Asl2003}. It was assumed that there exists a one to one correspondence between the branches of Lambert W function and the characteristics roots of the system. For scalar systems, the rightmost root corresponds to the principal branch which determines their stability \cite{Mori}. These results for scalar systems can not be extended for multi-variable cases. Therefore, stability analysis of higher order TDS is to an extent based on observations \cite{Ulsoy2007,Yi2006}. Based on these assumptions and observations several reckoning works have been derived \cite{Yi2010a,Yi2012,Yi2014}. In \cite{Rudy,Rudyconf}, it is claimed that in general there does not exist a one to one correspondence as said in \cite{Worldscientific}. A counter example is devised to disprove the one to one correspondence which was formulated by the proposers of the Lambert W function methodology. Exploiting these key points, a reverse engineering approach is presented for associating the whole eigenspectrum with only two real branches of the Lambert W function. \\

The aim of this paper is to extend the applicability of the method presented in \cite{Rudy} by
\begin{itemize}
\item generalizing it from second order to \textit{n}th order system. This is a complement for the existing method \cite{Rudy}.
\item defining the common canonical (CC) form of TDS.
\item characterizing a new class of time delay systems which can be transformed into the CC form using state variable transformation.
\item exploiting these results to synthesize a controller.
\end{itemize}
Rest of the paper is structured as follows. Section 2 reassess the preliminaries about Lambert W function. Section 3 describes the main results of the paper. Numerical illustration is shown in Section 4 and finally, section 5 concludes the paper.  
\section{The Lambert \emph{W} function}
The Lambert \emph{W} function $x=W_k(z)$, is a multi-valued complex function if it satisfies (\ref{eq1a}) 
\begin{equation}\label{eq1a}
xe^x=z,
\end{equation}
\begin{figure}[ht]
\centering
\includegraphics[height=2in,width=3in]{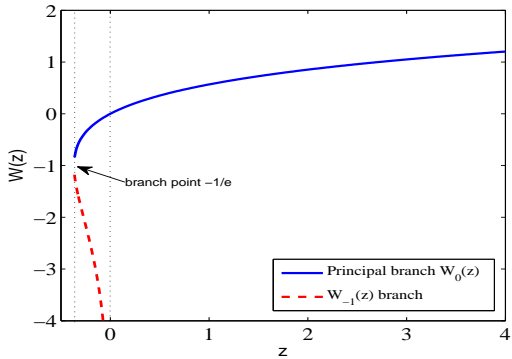} 
\caption{The two real branches of the Lambert W function.}
\label{realbranch}
\end{figure}
for $z\in\mathbb{C}$. It has infinite number of branches distinguished by a subscript $k=0,\pm1,\pm2,···,\pm\infty$, especially $W_0$ is called the principal branch. If $z\in\mathbb{R}$, then for $-1/e\leq z<0$,  $W(z)$ has two possible real values as shown in Fig. \ref{realbranch}. The branch satisfying $-1\leq W(z)$ is designated by $W_0(z)$ and the branch satisfying $W(z)\leq -1$ is denoted by $W_{-1}(z)$. The concept of branches has been discussed in \cite{Coreless} and hence not included here. 
\section{Main results}
In this section, we extend the applicability of the approach presented in \cite{Rudy} from second order to \textit{n}th order system. We propose a new class of systems which can be transformed into the CC form of TDS using state variable transformation to exploit these results. For this first we define the CC form of TDS.
\theoremstyle{definition}
\begin{defn}
A TDS (\ref{eq1}) is said to be in CC form if $A\in \mathbb{R^{\mathrm{n}\times \mathrm{n}}}$ is in companion form and $A_d\in \mathbb{R^{\mathrm{n}\times \mathrm{n}}}$ has all its rows zero except for the \textit{n}th row and is represented as
\begin{align}
\label{eq6}A=\begin{bmatrix}
0 & 1 & \hdots & 0\\
0 & 0 & \hdots & 0\\
\vdots & \vdots & \ddots & \vdots\\
a_{1} & a_{2} & \hdots & a_{n}
\end{bmatrix},
A_d = \begin{bmatrix}
0 & 0 & \hdots & 0\\
0 & 0 & \hdots & 0\\
\vdots & \vdots & \ddots & \vdots\\
a_{d1} & a_{d2} & \hdots & a_{dn}
\end{bmatrix}.
\end{align}
\end{defn}
In \cite{Rudy}, it is shown that the eigenspectrum of a second order TDS in the CC form, can be associated with only real branches of Lambert W function. These results have been extended for a \textit{n}th order system and stated in the form of a theorem. For this, we assume that (\ref{eq4}) is real and in companion form.
\newtheorem{theorem}{Theorem}
\begin{theorem}
The whole eigenspectrum of the system (\ref{eq1}) can be associated with only two real branches, $k=0$ and $k=-1$ of the Lambert \emph{W} function if it is in CC form.  
\end{theorem}
\vspace{-0.6cm}
\begin{pf}
Using the structure of $A_d$ in (\ref{eq6}), it is obvious that $M_k=hA_dP_k$, for any given $P_k$, has the form\vspace{-0.4cm}
\begin{equation}
	\label{eq7} M_k=\begin{bmatrix}
		0 & 0 & \hdots & 0\\
		\vdots & \vdots & \ddots & \vdots\\
		0 & 0 & \hdots & 0\\
		m_{1} & m_{2} & \hdots & m_{n}
	\end{bmatrix},
\end{equation}
where $m_{i}, i=1,2,\ldots,n$ are scalars. Based on the value of the element $m_n$ there are two possible cases:\\
\begin{figure*}
	\hrulefill
	\begin{align}
	\label{eq10}S_k&=\begin{bmatrix}
	0 & 1 & \hdots & 0 & 0\\
	\vdots & \vdots & \ddots & \vdots & \vdots\\
	0 & 0 & \hdots & 0 & 1\\
	\frac{m_{1}}{hm_{n}} W_k(m_{n})+a_{1} & \frac{m_{2}}{hm_{n}}W_k(m_{n})+a_{2} & \hdots & \frac{m_{n-1}}{hm_{n}}W_k(m_{n})+a_{n-1} & \frac{1}{h}W_k(m_{n})+a_{n}
	\end{bmatrix}.
	\end{align}
	\hrulefill
\end{figure*}
\textbf{Case 1: $m_{n}\neq 0$}\\
The matrix Lambert W function of $M_k$ is obtained as 
\vspace{-0.4cm}
\begin{align}
\label{eq8}W_k(M_k)&=\begin{bmatrix}
	0 & \hdots & 0 & 0\\
	\vdots & \ddots & \vdots & \vdots\\ 
	0 & \hdots & 0 & 0\\
	\frac{m_{1}}{m_{n}} W_k(m_{n}) & \hdots & \frac{m_{n-1}}{m_{n}} W_k(m_{n}) & W_k(m_{n})
	\end{bmatrix}
\end{align}
\begin{align}
\nonumber W_k(M_k)=\frac{W_k(m_{n})}{(m_{n})}M_k.
\end{align}
Therefore, matrix Lambert \emph{W} function of matrix $M_k$, is the $M_k$ matrix itself multiplied by a scalar constant $\frac{W_k(m_{n})}{h(m_{n})}$. Using (\ref{eq8}), $S_k$ is written as (\ref{eq9}), and given in (\ref{eq10}).
\begin{align}\label{eq9}
S_k=\frac{W_k(m_{n})}{h m_{n}}M_k+A,
\end{align}
\textbf{Case 2: $m_{n}=0$} \\
When $m_{n}=0$, then by using the following property 
\begin{align*} 
\nonumber \lim_{m_{n}\rightarrow 0} \frac{W_k(m_{n})}{m_{n}}; \quad  \lim_{x\rightarrow 0} \frac{x}{xe^x}=1
\end{align*}
we have $W_k(M_k)=M_k$. Hence, $S_k$ is given as
\begin{align*}
S_k=\begin{bmatrix}
0 & 1 & \hdots & 0 & 0\\
\vdots & \vdots & \ddots & \vdots & \vdots\\
0 & 0 & \hdots & 0 & 1\\
\frac{m_{1}}{h}+a_{1} & \frac{m_{2}}{h}+a_{2} & \hdots & \frac{m_{n-1}}{h}+a_{n-1} & \frac{1}{h}+a_{n}
\end{bmatrix}.
\end{align*}
Furthermore, the concept used here is to perform the steps given in Section 1, in reverse order to attribute branch index $k$. For this purpose we first formulate a real matrix $S_k$, which can be written in terms of its eigenvalues as 
\begin{align}
\squeezeup
\label{eq11}S_k&=\begin{bmatrix}
	0 & 1 & 0 & \hdots & 0\\
	\vdots & \vdots & \vdots & \ddots & \vdots\\
	0 & 0 & 0 & \hdots & 1\\
	w & x & y & \hdots & z
	\end{bmatrix},
\end{align}
where $w=(-1)^n\sum_{sets~of~n}{\lambda\lambda\lambda..\lambda},$\\$x=(-1)^{n-1}\sum_{sets~of~{n-1}}{\lambda\lambda..\lambda}$,\\$y=(-1)^k\sum_{sets~of~k}{\lambda\lambda..\lambda}$ and $z=-\sum_{sets~of~1}{\lambda}$ \cite{Brooks}. Comparing (\ref{eq10}) and (\ref{eq11}), yields\\
\squeezeup
\begin{align}
\label{eq12} W_k(m_{n})&=h(z-a_{n}),\\
	\nonumber :\\
		\label{eq13}
	m_{1}&=\frac{(w-a_{1})m_{n}}{(z-a_{n})}.
\end{align}
For real $S_k$, $M_k$ is real. Therefore, both sides of equation (\ref{eq12}) are real, which correspond to either $k=0$ or $k=-1$, depending on the scalar Lambert W function element $W_k(m_{n})$. For real arguments, the union of ranges of two real branches of the Lambert \emph{W} function that is $k=0$, the principal branch and $k=-1$, includes $\mathbb{R}$ \cite{Coreless}. \\
Further, it is necessary to show that $M_k$ is a solution of (\ref{eq5}), either for $k=0$ or $k=-1$. For this, let us assume that $v_1, v_2,...,v_n$ be the eigenvectors corresponding to $\lambda_1, \lambda_2, \ldots, \lambda_n$. The pair $(V, \Lambda)$ is an invariant pair of (\ref{eq1}), where
\begin{align}
	\nonumber V &=[v_1, v_2, \ldots, v_n]\\
	\label{eq14} \Lambda &=diag(\lambda_1, \lambda_2, \ldots, \lambda_n)
\end{align}
consequently, it must satisfy the characteristic equation
\begin{align}
	\label{eq15} \Lambda I-A-A_de^{-\Lambda h}=0.
\end{align}
multiplying by $V$ on both sides of (\ref{eq15}) 
\begin{align}\label{eq16}
	V\Lambda-VA-A_dVe^{-\Lambda h}=0
\end{align}
Noting that $S_k$ and $e^{-S_k}$ shares same set of eigenvectors, we have
\begin{align}
	\label{eq17} S_k=V\Lambda V^{-1}, \quad e^{-S_k h}=Ve^{-\Lambda h}V^{-1},
\end{align}
using (\ref{eq17}), it follows that
\begin{align}
	\label{eq18} V\Lambda=S_kV, \quad Ve^{-\Lambda h}=e^{-S_k h}V,
\end{align} 
substituting (\ref{eq18}) into (\ref{eq16}) yields
\begin{align}
	\label{eq19} [S_k-A-A_de^{-S_kh}]V=0.
\end{align}
since $V\neq0$. Hence
\begin{align}
	\label{eq20} S_k-A-A_de^{-S_kh}=0
\end{align}
substitute $S_k=\frac{1}{h}W_k(M_k)+A$ in (\ref{eq20}), yields in (\ref{eq5}). 
\qed
\end{pf}
\newtheorem{remark}{Remark}
\begin{remark}
	The above theorem is the extension of the approach used in \cite{Rudy}, from second order system to \textit{n}th order system. The applicability of this theorem is restricted to a certain class of systems, which are in the CC form. 
\end{remark}
Suppose, if a time delay system is not in CC form then the applicability of the Theorem 1 can be extended by means of the following theorem. For this, we characterize a new class of systems which can be transformed into the CC form using state variable transformation. 
\begin{theorem}\label{thm1}
Given a time delay system (\ref{eq1}) with $A_d = bc^T$, where  $b,c\in\mathbb{R}^{n\times 1}$, the system (\ref{eq1}) can be transformed into the CC form, if pair $(A,b)$ is controllable.
\end{theorem}
\begin{pf}
If we choose the structure $A_d=bc^{T}$, then (\ref{eq1}) is rewritten as
\begin{equation}\label{eq21}
\dot{x}(t)=Ax(t)+bc^{T}x(t-h).
\end{equation}
We assume that there exist a nonsingular state transformation matrix $T$, such that similarity transformation takes place. The change of variables is represented by a linear transformation
\begin{equation}\label{eq22}
x=Tz,
\end{equation} 
where $z$ is the state vector in the transformed domain. Transformation matrix $T$ is chosen as 
\begin{equation}\label{eq23}
T=UU^{-1}_c,
\end{equation}
where $U$ and $U_c$ are the controllability matrices of pair $(A,b)$ and ($\bar{A},\bar{b}$) respectively. ($\bar{A},\bar{b}$) is the controllable companion form of pair $(A,b)$ \cite{CTChen}. Substituting (\ref{eq22}) into (\ref{eq21}) obtains
\begin{equation}\label{eq24}
\dot{z}(t)=\bar{A}z(t)+\bar{A_d}z(t-h),
\end{equation}
where \begin{align}
\label{e25}\bar{A}&=T^{-1}AT
=\begin{bmatrix}
0 & 1 & 0 & \hdots & 0\\
0 & 0 & 1 & \hdots & 0\\
\vdots & \vdots & \vdots & \ddots & \vdots\\
* & * & * & \hdots & *
\end{bmatrix},
\\\nonumber~\\
\label{eq26}\bar{A_d}&=T^{-1}b c^{T}T, \quad \quad \bar{b}=T^{-1}b,\\
with \label{eq27}\quad &T^{-1}b=\begin{bmatrix}
0\\ \vdots \\ 1
\end{bmatrix}, \quad c^{T}T=\begin{bmatrix}
* & * & \hdots & *
\end{bmatrix}. 
\end{align}
where $*$ represents any value.
Finally, from (\ref{eq26}) and (\ref{eq27}), we observed that $\bar{A_d}$ has a structure $\begin{bmatrix}
0 & 0 & \hdots & 0\\
0 & 0 & \hdots & 0\\
\vdots & \vdots & \ddots & \vdots\\
* & * & \hdots & *
\end{bmatrix}$, which is the CC form of time delay systems. \qed
\end{pf}
\begin{remark}
The above analysis shows that after transforming the system (\ref{eq21}) into the CC form all the characteristics roots can be computed using only real branches of the Lambert W function corresponding to $k=0$ and $k=-1$. This is illustrated by an example in the subsequent section.
\end{remark}

The above results are used to design a stabilizing controller by assigning a subset of eigenvalues of the closed loop system in the subsequent section.
\subsection{Controller synthesis}
Controller design for time delay systems based on eigenvalue assignment has already been discussed in \cite{Yi2010a,Yi}, but in this approach, to find the auxiliary matrix $P$ is difficult and also, it is a hit and trial approach to find the controller $K$, with a specific set of initial conditions. These issues can be bypassed by the approach presented here.

Consider a time delay system with input delay
\begin{equation}\label{eq34}
\dot{x}(t)=Ax(t)+Bu(t-h),
\end{equation}  
with feedback control
\begin{equation}\label{eq35}
u(t)=Kx(t).
\end{equation}
The closed-loop system is written as
\begin{equation}\label{eq36}
\dot{x}(t)=Ax(t)+BKx(t-h).
\end{equation}
The solution matrix is written in terms of Lambert \emph{W} function as
\begin{equation}\label{eq37}
S_k=\frac{1}{h}W_k(M_k)+A.
\end{equation}
where $M_k=hBKP_k$. To assign the desired eigenvalues of TDS in left half of the complex plane, controller gain $K$ is obtained by the following algorithm
\begin{algorithm}
\caption{}\label{algo1}
\begin{algorithmic}[1]
\State Select desired characteristic roots $\lambda_{i,des}$ for $i=1,...,n$.
\State Create a matrix $S_k$ using (\ref{eq12}), so that it returns selected eigenvalues.
\State Obtain $W_k(M_k)=h(S_k-A)$ from (\ref{eq37}).
\State Compute $M_k$ from step 3.
\State Substitute $P_k=e^{-S_kh}e^{h(S_k-A)}$ in $M_k=hBKP_k$ and then compare both sides of it to obtain $K$.
\end{algorithmic}
\end{algorithm}
\begin{remark}
This approach fixes the $P_k$ matrix for a subset of eigenvalues, hence resolves the trouble of selecting appropriate auxiliary matrix. The controller gain is the only parameter to be found instead of $K$ and $P_k$ of the eigenvalue assignment method. Since it is a straight-forward method, therefore, no need of hit and trial using different sets of initial conditions to assign the selected eigenvalues at desired locations \cite{Yi2010a,Worldscientific}. 
\end{remark}
Using Algorithm 1, a subset of closed loop characteristic roots can be placed at desired locations. But it does not guarantee the overall stability of the system. In the literature, it is reported that the roots corresponding to the principal branch of Lambert W function gives the rightmost root which determines stability. However, recent study of \cite{Rudy} claims that for the systems in CC form several roots are associated with the principal branch and practically, it is hard enough to identify that which one is the rightmost among them. Therefore, the Lambert W function based method alone, is not well suited to guarantee the stability. Hence, in the present study, stability is investigated with the help of Nyquist plot which guarantees the stability by ensuring that the roots placed at desired locations using the approach presented above are rightmost one and is given in the following subsection. 
\subsection{Stability analysis using Nyquist plot}
System (\ref{eq1}) is stable, if all the roots of characteristic polynomial $p(\lambda)=\lambda^n-a_{n}\lambda^{n-1}-\ldots-a_{1}-a_{dn}\lambda^{n-1}e^{-\lambda h}-\ldots-a_{d1}e^{-\lambda h}$, have negative real parts. Stability of this system is investigated using the following definition which is based on the Nyquest stability criteria for time delay systems\cite{Wang}.
\begin{defn}
A linear-time-invariant system with delay is said to be asymptotically stable if and only if the Nyquist plot of 
\begin{equation}\label{eq38}
\frac{p(j\omega)}{(1+j\omega)^n},
\end{equation}
does not encircle the origin of the complex plane, where $n$ is the degree of $p(\lambda)$. Let $\lambda=\sigma+j\omega+\alpha$, then $Re(\lambda)\leq\alpha$ if and only if $\sigma\leq 0$. Therefore, $\lambda=\lambda^*$ is the rightmost root if it satisfies the following conditions
\begin{itemize}
\item The Nyquist plot of 
\begin{equation}\label{eq39}
\frac{p(j\omega+\lambda^*)}{(j\omega+1)^n},
\end{equation}
passes through the origin, and
\item The Nyquist plot of 
\begin{equation}\label{eq40}
\frac{p(j\omega+\lambda^*+\mu)}{(j\omega+1)^n},
\end{equation}
does not encounter the origin of the complex plane for every small $\mu>0$.
\end{itemize} 
\end{defn}
Therefore, If the Nyquist diagram does not encircle the origin, then it is confirmed that the rightmost root that guarantee the stability has been computed correctly. This approach is demonstrated in following illustrative examples.
\section{Numerical example}
\subsection{Analysing the roots}
\textbf{Example 1:} Consider a time delay system which is not in CC form, with matrices given as
\begin{align*}
A&=\begin{bmatrix}
-1 & 2 & -1\\
-4 & -1 & -3\\
-2 & -3 & -2\end{bmatrix}, \quad A_d = \begin{bmatrix}
1 & -1 & 2\\
0 & 0 & 0\\
-1 & 1 & -2\end{bmatrix}\\
b&=\begin{bmatrix}
-1\\
0\\
1
\end{bmatrix}, \quad c=\begin{bmatrix}
-1\\1\\-2
\end{bmatrix}^T, \quad h=2.
\end{align*}
The state variable transformation matrix $T$ is
\begin{align*}
T = UU_c^{-1}=\begin{bmatrix}
0 & -4 & -1\\
3 & 1 & 0\\
-1 & 4 & 1\end{bmatrix}.
\end{align*}
and the transformed system into the CC form is obtained as
\begin{align*}
&\bar{A}=\begin{bmatrix}
0 & 1 & 0\\
0 & 0 & 1\\
-7 & -2 & -4\end{bmatrix}, \quad\bar{A_d} = \bar{b}\bar{c}^T=\begin{bmatrix}
0 & 0 & 0\\
0 & 0 & 0\\
5 & -3 & -1\end{bmatrix}\\
&\bar{b}=\begin{bmatrix}
0\\
0\\
1
\end{bmatrix}, \quad \bar{c}=\begin{bmatrix}
5\\-3\\-1
\end{bmatrix}^T.
\end{align*}
In pursuance of obtaining an analytical estimate of $P_k$, we use \lq\lq reverse-engineering\rq\rq approach, as in the proof of Theorem 1. For this, first the roots of the system are obtained using QPmR algorithm \cite{Zitek}.

\begin{figure}
	\centering
	\includegraphics[height=2in,width=3in]{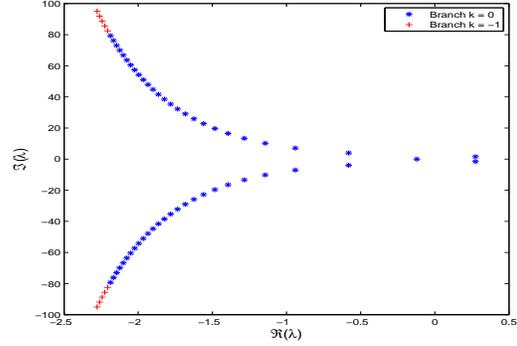} 
	\caption{Eigenspectrum of the system in Example 1.}
	\label{eigen}
\end{figure}

Let the dominant roots of the system be $\lambda_1 = -0.1211,~ \lambda_2 = 0.2744+1.5588i,~\lambda_3 = 0.2744-1.5588i$. Corresponding to these roots $S_k$ matrix is
\begin{align}
\nonumber S_k =\begin{bmatrix}
0 & 1 & 0\\
0 & 0 & 1\\
-0.3034 & -2.4386 & 0.4277
\end{bmatrix}.\\
\label{eq41}
W_k(M_k)=\begin{bmatrix}
0 & 0 & 0\\
0 & 0 & 0\\
13.3932 & -0.8772 & 8.8553
\end{bmatrix}.
\end{align}
From (\ref{eq41}) we have $W(m_{3})=8.8553\in[-1,\infty)$, which is the range of the principal branch of the Lambert W function. Therefore, there exists a matrix $M_k$ for which (\ref{eq41}) is satisfied for $k=0$, and that matrix is
\begin{align}\label{eq42}
M_0 = \begin{bmatrix}
0 & 0 & 0\\
0 & 0 & 0\\
9.3908 & -0.6151 & 6.2090
\end{bmatrix}\times 10^4.  
\end{align}
Since $A_d$ and $M_0$ are singular matrices, hence there are infinite number of $P_0$ matrices that satisfy (\ref{eq42})
for $k=0$. One of such matrices is
\begin{align*}
P_0 = \begin{bmatrix}
0.0001 & 0.0001 & 0.0001\\
0.0001 & 0.0001 & 0.0001\\
-4.6952 & 0.3077 & -3.1043
\end{bmatrix}\times 10^4.  
\end{align*}
The difficulty of making initial guess for auxiliary matrix $P_0$ has been resolved using this method. If this value of $P_0$ is taken as starting value while solving nonlinear equation (\ref{eq5}) using Lambert W function based method for $k=0$ then corresponding $S_k$ matrix and its eigenvalues are obtained after few iterations, and computation time is very small approximately less than $2$ or $3$ sec.\\
Now let us choose some other characteristic roots as $\lambda_1 = -0.1211,~ \lambda_2 = -0.9405+7.0675i,~\lambda_3 = -0.9405-7.0675i$.
Corresponding to these roots $S_k$ matrix is
\begin{align}
\nonumber S_k=\begin{bmatrix}
0 & 1 & 0\\
0 & 0 & 1\\
-6.1567 & -51.0613 & -2.0021
\end{bmatrix}.\\
\label{eq43}W_k(M_k)=\begin{bmatrix}
0 & 0 & 0\\
0 & 0 & 0\\
1.6867 & -98.1226 & 3.9957
\end{bmatrix}.
\end{align}
From this we have $W(m_{3})=3.9957\in[-1,\infty)$, which is the range of the $k=0$ branch. Therefore, matrix $M_0$ for which (\ref{eq43}) is satisfied for $k=0$, is
\begin{align}\label{eq44}
M_{0} = \begin{bmatrix}
0 & 0 & 0\\
0 & 0 & 0\\
0.0917 & -5.3345 & 0.2172
\end{bmatrix}\times 10^3.  
\end{align}
using this $P_{0}$ matrix is obtained as
\begin{align*}
P_0 = \begin{bmatrix}
0.001 & 0.001 & 0.001\\
0.001 & 0.001 & 0.001\\
-0.0438 & 2.6693 & -0.1066
\end{bmatrix}\times 10^3 .   
\end{align*}
which is a solution to (\ref{eq5}) for $k =0$. Choosing initial conditions close to this matrix assures convergence to this solution.
Fig. \ref{eigen} shows some characteristic roots of the system in Example 1, computed using $k=0$ \& $k=-1$ branches.
\\
Thus it is shown that how the whole eigenspectrum of a class of systems with the structure specified in (\ref{eq21}) is analysed using only two branches of the Lambert W function and the appropriate initial conditions for the solution of the nonlinear equation (\ref{eq5}) are obtained.
\subsection{Controller synthesis}
\begin{figure}[ht]
	\centering
	\includegraphics[height=2in,width=3in]{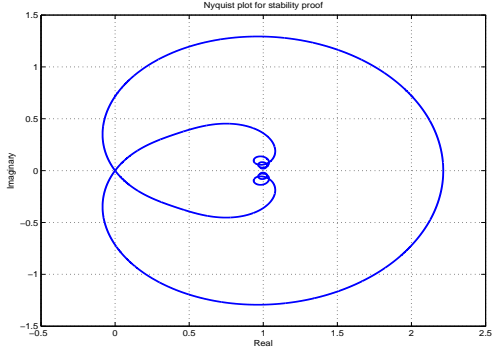} 
	\caption{Proof of stabilization of the system by Nyquist plot of $\Delta(jw-1)/(1+jw)^2$ which passes through origin of the complex plane.}
	\label{nyquist}
\end{figure}

\begin{figure}[ht]
	\centering
	\includegraphics[height=2in,width=3in]{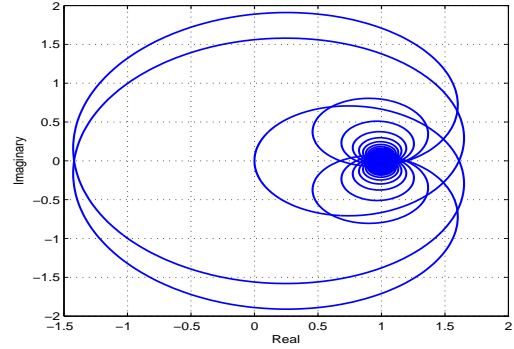} 
	\caption{Proof of stabilization of the system by Nyquist plot of $\Delta(jw-1)/(1+jw)^3$ which passes through origin of the complex plane.}
	\label{nyquist2}
\end{figure}

\textbf{Example 2:} Consider the van der Pol equation with system matrices \cite{Yi2010a}
\begin{align} \label{eq45}
&A=\begin{bmatrix}
0 & 1\\
-1 & 0.1
\end{bmatrix}, \quad b=\begin{bmatrix}
0\\
1
\end{bmatrix}, \quad h=0.2.
\end{align}
Without delayed feedback term this system is unstable because its rightmost eigenvalues ($\lambda=0.05\pm0.9987i$) lie in the right of the complex plane. For stability, let us choose the desired eigenvalues be $-1\pm2i$. The controller gain is found by using Algorithm \ref{algo1}. $S_k$ and corresponding $W_k(M_k)$ is are obtained as
\begin{align*}
S_k=\begin{bmatrix}
0 & 1\\
-5 & -2 
\end{bmatrix}, \quad
W_k(M_k)=\begin{bmatrix}
0 & 1\\
-0.8 & -0.42 
\end{bmatrix}.
\end{align*}
Since $W(m_{2})=-0.42 \in [-1,-\infty)$ hence $k=0$. Further $M_0$ and $P_0$ matrices are
\begin{align*}
M_0&=\begin{bmatrix}
0 & 1\\
-0.5256 & -0.2760 
\end{bmatrix}.\\
P_0&=\begin{bmatrix}
1.0425 & -0.1563\\
0.2988 & 0.8954
\end{bmatrix}.
\end{align*} 
Then, the required controller is $K=[-1.9802 ~ -1.8865]$.\\
Finally, it remains to show that the desired characteristic equation roots are the rightmost one ensuring stability of the system. The Nyquist plot is considered for this purpose. The Nyquist plot of $\Delta(jw-1)/(1+jw)^2$ shown in Fig. \ref{nyquist}, passes through the origin and confirms the stability of the system by ensuring that the desired characteristics roots are the rightmost.
\begin{figure}[ht]
	\centering
	\includegraphics[height=2in,width=3in]{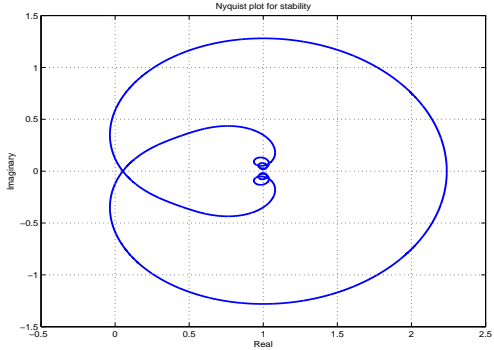} 
	\caption{Proof of stabilization of the system by Nyquist plot of $\Delta(jw-1+0.1)/(1+jw)^2$, which does not encircle the origin of the complex plane.}
	\label{nyquist1}
\end{figure}\\
\textbf{Example 3:} Recall the system in Example 1 with matrices
\begin{align*}
&A=\begin{bmatrix}
0 & 1 & 0\\
0 & 0 & 1\\
-7 & -2 & -4\end{bmatrix}, \quad A_d = \begin{bmatrix}
0 & 0 & 0\\
0 & 0 & 0\\
5 & -3 & -1\end{bmatrix},\quad b=\begin{bmatrix}
0\\
0\\
1
\end{bmatrix}.
\end{align*}
The rightmost eigenvalues of this system are $0.2744\pm1.5588i$, that is in the right half of the complex plane. Hence this system is unstable. To make this system stable, let us choose a subset of closed loop characteristic roots say $\lambda_1=-1, \lambda_2=-2$ and $\lambda_3=-3$. The $S_k$ and $W_k(M_k)$ matrices are
\begin{align*}
S_k=\begin{bmatrix}
0 & 1 & 0\\
0 & 0 & 1\\
-6 & -11 & -6
\end{bmatrix}, \quad W_k(M_k)=\begin{bmatrix}
0 & 0 & 0\\
0 & 0 & 0\\
2 & -18 & -4
\end{bmatrix}.\end{align*}
Since $W_k(3)=-4\in(-\infty,-1]$, hence $k=-1$. Corresponding to this matrices $M_{-1}$ and $P_{-1}$ are
\begin{align*}
M_{-1}&=\begin{bmatrix}
0 & 0 & 0\\
0 & 0 & 0\\
0.0366 & -0.3297 & -0.0733
\end{bmatrix}.\\
P_{-1}&=\begin{bmatrix}
0.3358 & -0.2610 & 0.0028\\
-1.1501 & 0.8101 & -0.0092\\
3.7836 & -2.4802 & 0.0293
\end{bmatrix}\times10^3.
\end{align*}
The controller is obtained as $K_d=[-2.3316 ~ 4.9380 ~  1.3523]$. Nyquist plot of $\Delta(jw-1)/(1+jw)^3$ shown in Fig. \ref{nyquist2}, does not encircle the origin and confirms the stability of the system ensuring that the desired characteristics roots are the rightmost.
\\
\begin{figure}[ht]
	\centering
	\includegraphics[height=2in,width=3in]{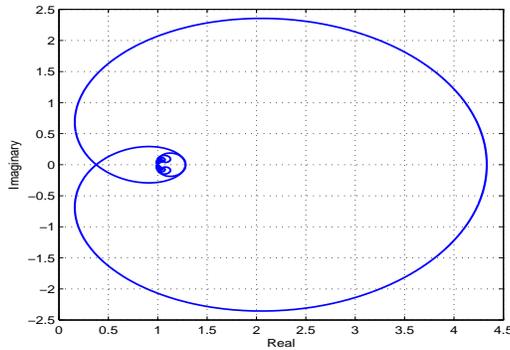} 
	\caption{Proof of stabilization of the system by Nyquist plot of $\Delta(jw)/(1+jw)^3$, which does not encircle the origin of the complex plane.}
	\label{nyquist2a}
\end{figure}
\section{Conclusion}
In this work, the attribution of branch number to eigenspectrum is generalized to include a class of \textit{n}th order TDS. These results extend the results of \cite{Rudy} which were valid only for second order systems. We characterized a class of TDS which can be transformed into the CC form using a state variable transformation. The characteristic roots of the proposed class of TDS can be analysed using only real branches of the Lambert W function. Moreover, the obtained results are utilized to synthesize a controller for placing a subset of eigenvalues at desired locations. Stability is analysed with the help of the Nyquist plot. It is shown that there are many roots which correspond to the principal branch $k=0$ and $k=-1$ branch. Among these several roots, it is difficult to identify the rightmost, that determines stability of the system, and is a topic for further research.

\end{document}